\documentclass[twocolumn,showpacs,preprintnumbers,amsmath,amssymb]{revtex4}
\usepackage{natbib}

\usepackage{amsmath,amsthm,amsfonts,graphicx,epsfig}
\usepackage[usenames]{color}

\begin{document}
\title{Vectorial Phase Retrieval for Linear Characterization of Attosecond Pulses}
\author{O. Raz$^{1}$, O. Schwartz$^{1}$, D. Austin$^{2,3}$, A. S. Wyatt$^{3}$, A. Schiavi$^{3}$, O. Smirnova$^{4}$, B. Nadler$^{1}$, I. A. Walmsley$^{3}$, D. Oron$^{1}$, N. Dudovich$^{1}$}
\affiliation{$^{1}$Weizmann Institute of Science, 76100, Rehovot, Israel\\$^{2}$ICFO Parc Mediterrani de la Tecnologia
,08860, Barcelona, Spain\\$^{3}$Department of
Physics,University of Oxford, Clarendon Laboratory, Oxford, UK \\
$^{4}$Max-Born Institute for Nonlinear Optics and Short Pulse Spectroscopy, D-12489 Berlin, Germany }

\email{feraz@weizmann.ac.il}

\date{\today}%
\begin{abstract}

The waveforms of attosecond pulses produced by high-harmonic generation carry information on the electronic structure and dynamics in atomic and molecular systems. Current methods for the temporal characterization of such pulses have limited sensitivity and impose significant experimental complexity. We propose a new linear and all-optical method inspired by widely-used multi-dimensional phase retrieval algorithms. Our new scheme is based on the spectral measurement of two attosecond sources and their interference. As an example, we focus on the case of spectral polarization measurements of attosecond pulses, relying on their most fundamental property -- being well confined in time. We demonstrate this method numerically reconstructing the temporal profiles of attosecond pulses generated from aligned $CO_2$ molecules.
\end{abstract}
\maketitle     
Optical and XUV pulses with durations significantly below one picosecond cannot be directly characterized in the time domain, since there are no suitable photodetectors. Therefore they are usually characterized in the frequency domain by measuring both  the spectral amplitude and spectral phase of the pulse.  The former may be determined straightforwardly by means of a spectrometer. The latter requires either a fast modulator, a detector or a reference pulse with which the unknown pulse can be  interfered. A modulator or gate of sufficiently rapid response is typically synthesized by means of nonlinear optical processes \cite{NoLinearPhase}. In the femtosecond regime, most measurement schemes are based on nonlinear light matter
interactions. By exploiting media having a nonlinear response, various successful schemes have been developed for complete characterization of femtosecond pulses, most notably FROG \cite{FROG} and SPIDER \cite{SPIDER}.

The ability to produce attosecond pulses has set new benchmarks in time-resolved measurements. Such pulses enable one to probe electron dynamics on the atomic time scale. Recently it has been demonstrated that the attosecond production process carries information about both electron dynamics \cite{OlgaNature} and molecular structure \cite{Tomography}. However, full access to the information contained in the pulse requires its complete characterization. A direct implementation of the pulse characterization schemes developed for the femtosecond regime is challenging, due to the low signal levels and the absence of appropriate nonlinear media in the XUV.

Several characterization schemes have been developed for the attosecond XUV domain \cite{CRAB-FROG,FROGIonization,RABBIT2,StreakCamera,WalmsleyAtto,JapanCharacterization}. However, time resolved measurement of attosecond pulses remains a major challenge. For example, certain important attosecond-scale processes such as plasma mirrors \cite{Plasma_Mirrors} and field-enhancement HHG \cite{FieldEnhancement_HHG} have not yet been fully characterized.

In this letter we propose a new approach for measuring attosecond pulses, which - in contrast to other methods, relies only on
\emph{linear spectral measurements} of the radiation generated by several sources.
The key new feature of the method that enables this approach is the explicit utilization of a temporal support constraint in the retrieval algorithm. The fact that the pulse is limited in duration allows a unique solution to be extracted.
To illustrate the approach, we consider the two polarization components of the attosecond pulse as two independent sources. We show that the spectral measurement of the two polarizations together with their relative phases, obtained by means of spectral interference, is sufficient to retrieve the spectral phase of each component \textit{individually}.
We numerically test the method on attosecond pulses calculated for aligned $CO_2$ molecules \cite{Olga}, in which the polarization varies non-trivially with the frequency. We establish excellent reconstructions even for pulses that do not exactly satisfy the finite duration assumption, but rather have a sufficiently rapid decay.


Phase retrieval problems are common in many branches of physics, including astronomy \cite{FienupBook}, NMR \cite{NMR_PhaseRet}, ultra-fast optics \cite{FROG}, crystallography \cite{miao2008extending} and lens-less imaging \cite{LenslessPRL,Lensless_NatPhys_2006,NatureLenslessJila}. In each of these cases, the measured signal spectrum, together with some assumption on the signal itself, allows one to retrieve the phase, without directly measuring it. In our case, the assumed property is that the pulse has a limited, but not necessarily  known, ``time window'' in which the intensity is non-zero, usually referred to as \emph{compact support}. Generic compact support phase retrieval problems are known to have many solutions in the one-dimensional ($1D$) case, but usually a unique solution for higher dimensions \cite{Sodin}. As our measurements are of a one dimensional Fourier transform, the compact support assumption is not enough to reconstruct the spectral phase. This is solved by measuring the \emph{relative} phase between two (or more) spectra.
Such a measurement enables a \emph{vectorial} $1D$ phase retrieval problem. If the two components of the vector are non-degenerate (in a sense to be defined later on),  then the problem has a unique solution for the spectral phase, up to a phase linear in frequency (i.e. a arbitrary delay which is not physically significant). It is useful to consider how the temporal support constraint enables a unique inversion of the data. We first develop a formal approach that illustrates this, and then describe a simpler inversion algorithm.


Consider the electric field spectrum $E(\omega)$ of a pulse with a finite duration $T$, sampled\footnote{The continuous case will be discussed in a future publication} at $\omega_j=2\pi/T$ for $j=1,...,N$. The spectrum can be written as $E(\omega)=\sum_{t}{\hat E(t)e^{-i\omega t}}=\sum_{t}{\hat E(t)z^t}$, where $z=e^{-i\omega}$. According to the fundamental theorem of algebra, we can write  $E(z)=\sum_{t}{E(t)z^t}=\prod_{j}{(z-z_j)}$ where $z_j$ are the $N$ roots of the polynomial $E(z)$. A linear measurement of the spectrum measures $|E(\omega)|^2$, which at the sampled points $|z| = 1$ can be shown to equal:
\begin{equation}\label{Basic_Ztrans}
|E(z)|^2=\frac{\prod{\overline{z_j}}}{(-z)^N}\prod_{j}{(z-z_j)(z-\overline{z_j}^{-1})}
\end{equation}
where  $\overline{z_j}$ is the complex conjugate of $z_j$. Without
any prior knowledge about the pulse, the spectral phase is, by
definition, unrecoverable. This is because the polynomial representing the spectrum is under-sampled: $E(\omega)$ corresponds to a polynomial of degree $N$ in $z$, but $|E(\omega)|^2$ corresponds to a polynomial of degree $2N$ in $z$. Therefore, the $N$ samples of $|E(z)|^2$ are insufficient to unambiguously determine $\hat E(t)$. If,
however, a compact support constraint is assumed, meaning $\hat E(t)=0$
for, say the $N/2$ samples in the range $T/ 2\leq t\leq T$, the degree of $|E(z)|^2$ is at most $N$. Therefore, it is well sampled and the $N$ roots of the polynomial ($z_j$ and $\overline{z_j}^{-1}$) are uniquely determined. The compact
support constraint, nevertheless, does not remove all the
ambiguities. In order to retrieve the pulse $E(z)$, only one root
from each pair of roots, $z_j$ and $\overline{z_j}^{-1}$, should be
chosen. As there are $2^{N/2}$ possibilities to choose one root from
each pairs of $|E(z)^2|$, there still exist $2^{N/2}$ different
pulses, all having both the measured spectrum and the correct
compact support. This ambiguity in the $1D$ phase retrieval problem
is well known \cite{Sodin}. Hence, additional information is
required to unequivocally determine the ``correct'' choice.

\begin{figure}
  \includegraphics[width=10cm]{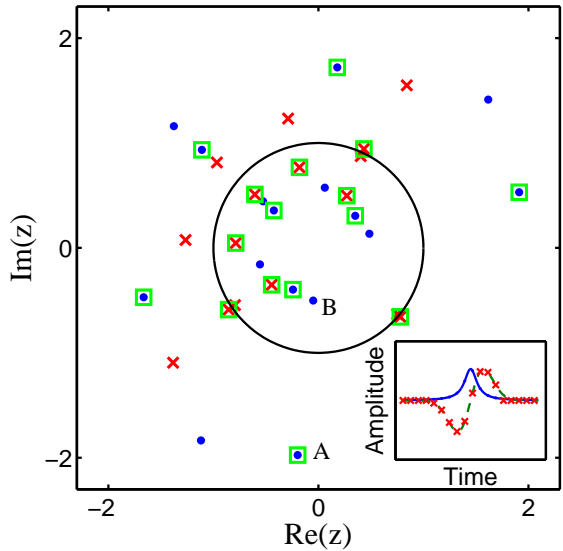}\\
  \caption{An example for our reconstruction procedure for two artificially generated pulses with the same compact support. The blue dots, red crosses and green squares are the roots, in the complex plane, of $|E_x(z)|^2$,$|E_y(z)|^2$ and $E_x(z)\overline{E_y(z)}$ respectively. A time domain plot of the two polarizations is shown in the inset. }\label{Fig:RootsRec}
\end{figure}

The phase ambiguity can be overcome by using two (or more) spectra which we refer to as ``components'', and the \emph{relative phase} between them. We will denote such measurements as ``vectorial''. As an example of a vectorial measurement, appropriate for the attosecond domain, we use polarization: when attosecond pulses are generated from an anisotropic media,
such as aligned molecules by means of high-harmonic generation, non-trivial frequency-dependent polarization is expected \cite{Jerom_Polarization_InHHG}. For the two component spectra, $|E_{x}(\omega)|^2$ and $|E_{y}(\omega)|^2$
with the same time domain compact support, we can find the $N$ roots
$\{z_{x}\}_j$, $\{z_y\}_j$ corresponding to Eq.~(\ref{Basic_Ztrans})
and their complex conjugates. Without the relative phase, these are two independent $1D$ phase retrieval problems, each having many solutions. However, a complete polarization measurement, consisting of both the spectra of two orthogonal polarizations as well as the interference spectrum between them, provides sufficient information to eliminate the ambiguities. Noting that
$E_{x,y}(\omega)=|E_{x,y}(\omega)|e^{\phi_{x,y}(\omega)}$, the
vectorial measurements allows us to establish
$|E_x(\omega)|^2,|E_y(\omega)|^2$ and
$E_x(\omega)\overline{E_y(\omega)}={|E_x||E_y|}e^{i(\phi_x-\phi_y)}$.
These quantities are represented by the factored polynomials:
\begin{eqnarray}
|E_x(z)|^2=\frac{\prod{\overline{z^x_j}}}{(-z)^N}\prod_{j}{(z-z^x_j)(z-\overline{z^x_j}^{-1})}\label{Eq:RootsX} \\
|E_y(z)|^2=\frac{\prod{\overline{z^y_j}}}{(-z)^N}\prod_{j}{(z-z^y_j)(z-\overline{z^y_j}^{-1})}\label{Eq:RootsY} \\
E_x(z)\overline{E_y(z)}=\frac{\prod{\overline{z^y_j}}}{(-z)^N}\prod_{j}{(z-z^x_{j})(z-\overline{z^y_{j}}^{-1})} \label{CrossTerm}
\end{eqnarray}
The essential point is that from these equations the ``which root''
ambiguities of both components can be resolved: the correct roots
for $E_{x}(z)$ are those that are common to $|E_{x}(z)|^2$
and $E_x(z)\overline{E_y(z)}$, and similarly for $E_y(z)$. In
Fig.\ref{Fig:RootsRec}, we show an example of the method for a
simple pulse. The time domain signals are shown in the inset. In the
main figure, the roots generated by
Eqs.~(\ref{Eq:RootsX},\ref{Eq:RootsY}) are shown as dots and crosses
in the complex plane (the black line is the unit circle). As
expected, the roots come in pairs: for example, the roots marked by
$A$ and $B$ are related by $z_A=\overline{z_B}^{-1}$. The one
dimensional phase retrieval problem requires one to choose the
correct root from each pair. In the vectorial case, this can be done
using the roots generated by Eq.(\ref{CrossTerm}): these roots are
marked by green squares. As seen in the figure, each square
coincides with only one of the roots, thus identifying the
correct root of each pair. In the above example, $A$, rather than
$B$, is the correct root, as it coincides with a root of
Eq.~(\ref{CrossTerm}).

Clearly, if $|E_x(z)|^2$ and
$|E_y(z)^2|$ have $m$ common zeros, there is still an ambiguity among $2^m$ different pulse shapes.
We will call such cases \emph{degenerate}. In the absence of noise, the set of pulses which share common zeros is very small, unless they have been manipulated in a common manner to introduce them. Nevertheless, common zeros do arise when both pulses are passed through the same linear phase-only filter. The pulses then share a component of common spectral phase, leading to degeneracy. In this case the pulse fields cannot be retrieved by the above method. Further, most experimental data will have some noise $S(z)$. This will tend to move the zeroes of the polynomials from $z_i$ to $z_i - \Delta z_i$. A simple calculation shows that $$E(z) + S(z) =E(z) + \sum_i{\Delta z_i\prod_{j\neq i}{(z-z_j)}} + \mathbb{O}(\Delta z^2)$$ Assuming that all the zeros are shifted by about the same distance , the position uncertainty radius scales with $N$ as $\Delta z\sim1/N$. In contrast, the distance between the zeros scales with $N$ as $1/\sqrt{N}$. Hence, increasing the spectral bandwidth (and therefore increasing $N$ without changing the compact support) allows the zeros to be completely distinguished.

%
%

Another consideration is that most physical pulses do not have exact compact support, but rather a sharp decay outside some temporal region. Our method is applicable even in such cases, as we now show by means of an alternative solution method.
$\hat E_{x}(t)$ and $\hat E_{y}(t)$ can be written as:
\begin{eqnarray}
\hat E_x(t) = \sum_{\omega}{E_x(\omega)e^{i\omega t}} = \sum_{\omega}{|E_x(\omega)|e^{i\omega t}X(\omega)}\label{Eq:Linear}\\
\hat E_y(t)  = \sum_{\omega}{|E_y(\omega)|e^{i\omega t}e^{i(\phi_y(\omega)-\phi_x(\omega))}X(\omega)}\label{Eq:Linear_Pol}
\end{eqnarray}
where  $X(\omega) = e^{i\phi_x(\omega)}$ represents the spectral
phase. Applying the compact support assumption gives
$\hat E_{x}(t)|_{t=1,...,\frac{N}{2}}=\hat E_{y}(t)|_{t=1,...,\frac{N}{2}}=0$. Using this condition in
Eq.~(\ref{Eq:Linear},\ref{Eq:Linear_Pol}) gives a set of $N$ linear
homogenous equations for the $N$ unknowns $X(\omega)$. 
Exploiting the
arbitrariness of the absolute phase, one can set  $X(\omega_1)=1$
and obtain an over-determined set of inhomogeneous
equations. By solving this set, one can solve the phase problem provided $X(\omega)$ is a phase-only function, that is, $|X(\omega)|=1$ for all $\omega$. This has two important consequences: (1) Since the problem boils down to solving linear equations, the sensitivity to noise is linear in the noise amplitude; (2) In cases where the solution to Eq.~(\ref{Eq:Linear},\ref{Eq:Linear_Pol}) is far from yielding a unimodular complex number, the compact support assumption is either wrong
or there is a degeneracy. As we later show, this gives us the means to search for the correct compact support without assuming it
beforehand.

\begin{figure}
  \includegraphics[width=8cm]{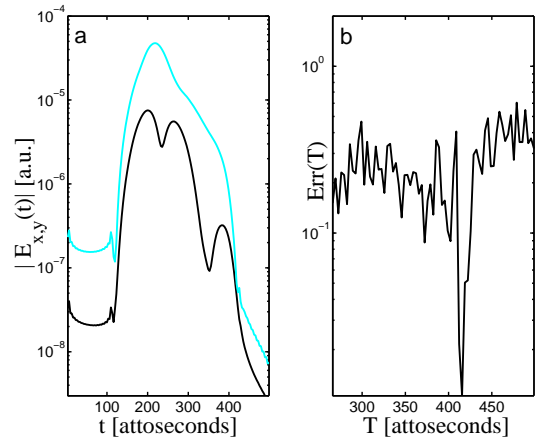}\\
  \caption{\textbf{(a)} Two time domain polarization components of the attosecond pulse generated from $CO_2$ molecules at an alignment of $58^o$ (between the molecule symmetry axis and the IR polarization) and IR laser intensity of $0.07 au$ at $800nm$. \textbf{(b)} Relative change in the pulse energy ( defined in Eq.\ref{Eq:GSMetric}) for different assumed compact supports.}\label{Fig:PulseInLog+CompSupScan}
\end{figure}

\begin{figure}
  \includegraphics[width=8cm]{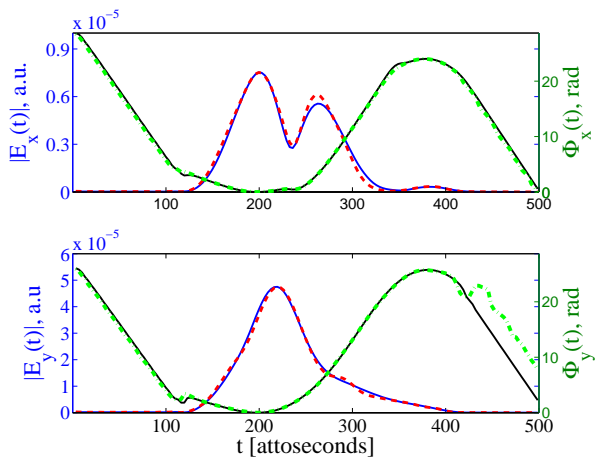}\\
  \caption{Comparison between original (Amplitude - blue, Phase - black) and retrieved (Amplitude -dashed red, Phase -- dash-dot green) pulses, for the two polarizations of the HHG pulse generated from the $CO_2$ molecule at an alignment of $58^o$ (between the molecule symmetry axis and the IR polarization) and IR laser intensity of $0.07 au$ at $800nm$. The lower figure shows the radiation in the IR polarization, and the upper shows the perpendicular polarization. The assumed pulse duration is 415 attoseconds. }\label{Fig:Pulses}
\end{figure}

Our proposed algorithm seeks a domain of compact support choosing that which is most consistent with the above conditions as our constraint. For each assumed domain of support, $T$, we resample the spectral information at discrete frequencies with spacing $\Omega = 2\pi/T$. We then use
$\hat E_{x,y}(t)|_{t=1,...,\frac{N}{2}}=0$ in Eqs.~(\ref{Eq:Linear}) and
(\ref{Eq:Linear_Pol}) to find $X(\omega)$. For each $T$ we
calculate how far $X(\omega)$ are from being unimodular complex, by
calculating the relative change in the pulse's energy when using $X(\omega)$ (which might not be unimodular) as the spectral
phase:
\begin{equation}\label{Eq:GSMetric}
Err(T)=\frac{\sum_{\omega}{|E(\omega)|^2|(1-X(\omega))|^2}}{\sum_{\omega}{|E(\omega)| ^2}}
\end{equation}
$Err(T)$ is the standard metric used in the Gerschberg-Saxton
algorithm \cite{FienupBook} for compact support phase retrieval
problems. A plot of $Err(T)$ is given in
Fig.\ref{Fig:PulseInLog+CompSupScan}. We then choose the value of $T$ that minimizes $Err(T)$ as the compact support domain and the argument of the corresponding $X(\omega)$ as the pulse's phase.
We demonstrate our method by reconstructing simulated attosecond pulses generated from aligned $CO_2$ molecules (for HHG simulation details see \cite{Olga}). The simulated attosecond pulses do not have compact support, but exhibit a sharp Gaussian decay (See Fig.\ref{Fig:PulseInLog+CompSupScan}, where the two original pulse polarizations are plotted in a logarithmic scale). Figure \ref{Fig:Pulses} shows the temporal profiles of the two polarizations of the simulated pulse and their reconstruction. As can be seen, their agreement is excellent. The errors are typically of the order of $Err(T)\sim0.05$.

There are many experimental ways to implement the method developed here. So far we have discussed the spectrally dependent polarization components of HHG which can be measured, using an XUV polarizer, up to a sign ambiguity of the phase difference. The sign ambiguity can be resolved by means of a wave plate. For many cases this is not needed as by continuity the ambiguity is the same for all $\omega$, which means an overall time direction ambiguity. Another way to use the same idea is to generate two XUV pulses from different sources, either spatially \cite{TwoSources,TwoSorcesOlga} or by using mixed gases \cite{MixedGasses}. By measuring the spectrum of each source alone and the interference between them, one can use our method to reconstruct the spectral phase. The spatial case can be generalized to spatio-temporal measurements by lateral shearing interferometry \cite{LateralShearing}.

To conclude, we have proposed and demonstrated a novel characterization method for attosecond pulses using a vectorial phase retrieval algorithm. Our method presents a new class of solutions of phase retrieval problems, applicable to many other fields, such as lensless imaging and optical spectroscopy. The main strength of the method lies in the fact that it removes the requirement for a nonlinear interaction which is currently a major limiting factor in our ability to resolve many attosecond processes. Extending our approach to characterize more complex electron dynamics proposes a new scheme of time resolved measurements where attosecond-scale phenomena can be observed using linear, time-stationary, apparatus.



\color{black}

\paragraph{ Acknowledgment} The authors would like to acknowledge financial support by the Israeli Ministry of Science Tashtiyyot program, the Crown center of photonics, the Minerva foundation and the ISF. O.R. acknowledge support by the Converging technologies fellowship of the Israeli Ministry of Science. O.S. is supported by the Adams Fellowship Program of the Israel Academy of Science and Humanities. IAW, AW, MM, AS acknowledge support from the UK EPSRC ( through grants EP/H000178/1, EP/F034601/1, and EP/E028063/1), the European Commission, through the ITN FASTQUAST, and the Royal Society.


\end{document}